\documentstyle[12pt]{article}

\newcommand{\beq}{\begin{equation}}
\newcommand{\eeq}{\end{equation}}
\newcommand{\la}{\langle}
\newcommand{\ra}{\rangle}
\begin{document}
\title{Form Factor $g$ In Longitudinal Space Charge Impedance}

\author{R. Baartman}
\date{June 1992}

\maketitle
\begin{abstract} 
In carrying out calculations of the effect of longitudinal space charge
on longitudinal motion, the transverse beam size appears in a form
factor which is usually written as $g=1+2\ln (b/a)$.  In fact, this
expression applies to particles with vanishing betatron amplitude in a
beam with uniform transverse distribution.  It is argued that an average
over the transverse distribution should be used instead of the value on
axis.  It is shown that for the realistic `binomial' family of
distributions the 1 in the above expression for $g$ should be replaced
by a value near 0.5 if $a$ is interpreted as twice the rms width of the
beam. 
\end{abstract}

\section{Introduction}

In an ideal machine with no space charge, the three degrees of freedom
of motion can be investigated separately.  With space charge, this is no
longer possible:  the longitudinal force on a particle depends upon its
transverse position in the bunch and vice versa.  It is still profitable
to some extent to study longitudinal and transverse motions separately,
but one must keep in mind that an approximation is being made.  It has
become standard practice to calculate longitudinal space charge effects
using the on-axis force because it was felt that this corresponds to a
`worst case'.  However, there is no basis for this approach unless one
is interested in the absolute longitudinal space charge limit (even in
that case, the `on-axis' approach is debatable). 

Since synchrotron motion is slow compared with betatron motion, it is
clear that in longitudinal simulations the longitudinal force should be
averaged over transverse positions.  (Conversely, in transverse
simulations, the transverse force should be taken as that corresponding
to an instantaneous longitudinal position.)  This note is devoted to
deriving a formula for the longitudinal force as averaged over
transverse distributions for the binomial family of distributions.  The
beam is assumed to be round (radius=$a$) in a concentric round beam
pipe (radius=$b$).  It is found that the usual transverse geometric
form factor ($g$), which is $1+2\ln (b/a)$ for the on-axis case of the
uniform distribution, can be written as $K+2\ln (b/a)$, where $K$
depends upon the distribution and the definition of $a$. To be fair it
should be pointed out that the error incurred by these assumptions
(neither the beam nor the beam pipe is round nor of constant dimension)
may very well be larger than the error incurred by using the on-axis
formula. 

\section{The Distribution}
The assumed distribution is 
\beq
\rho=(\mu+1){\lambda\over\pi a^2}\left(1-r^2/a^2\right)^\mu
\eeq
for $r<a$ and $\rho=0$ otherwise.  The normalization is 
$\int\!\!\int\rho dxdy=2\pi\int\rho rdr=\lambda$, the line density.  For 
$\mu=0$ this is the Kapchinsky-Vladimirsky distribution (round case).  
For $\mu\rightarrow\infty$ it is gaussian provided 
$a\rightarrow\infty$ as well.  This can be seen by finding the standard 
deviation.
\beq
\overline{r^2}={\int\rho r^2rdrd\theta\over\int\rho rdrd\theta}=
{2\pi\over\lambda}\int_0^a\rho r^3dr={a^2\over\mu+2}.
\eeq
Since $\overline{r^2}=\overline{x^2+y^2}=\overline{x^2}+\overline{y^2}=
2\overline{x^2}\equiv 2\sigma_x^2$, this can be written as
\beq
a=\sqrt{2(\mu+2)}\,\sigma_x.
\eeq
In the Sacherer-Lapostolle convention, the beam size is characterized 
not by $\sigma_x$ but by $\tilde{x}\equiv 2\sigma_x$.  This is handy 
because for the K-V case $\tilde{x}=a$.  For the general case,
\beq
\tilde{x}=\sqrt{2\over\mu+2}\: a.
\eeq

\section{The On-Axis Potential}
From Gauss' law, the electric field is radial with 
magnitude\footnote{Strictly speaking this is only correct if $\lambda$
is constant.  However it is a good approximation if, as is usually the 
case with proton machines, the length scale of the longitudinal 
variation is large compared with the beam pipe size, i.e.\ bunch length
$\gg b$.} 
\beq
E(r)={1\over \epsilon_0r}\int_0^r\rho (r')r'dr'=
{\lambda\over 2\pi\epsilon_0r}\left\{ \begin{array}{cl}
        1-\left(1-r^2/a^2\right)^{\mu+1} & \mbox{if $r<a$} \\
                      1                  & \mbox{otherwise} 
\end{array}\right..
\eeq

With the beam pipe (radius=$b$) at ground, the potential as a function 
of radius is
\beq
V=-\int_r^bE(r')dr'.
\eeq
Assuming all the beam is in the pipe ($a<b$), the potential in the beam
is 
\beq
V={-\lambda\over 2\pi\epsilon_0}
\left\{ \int_r^a\left[1-\left(1-r^2/a^2\right)^{\mu+1}\right]{dr\over r}
+\ln\left({b\over a}\right)\right\},
\eeq
which for $\mu$=integer can be expanded as 
\beq
V={-\lambda\over 4\pi\epsilon_0}
\left[\sum_{k=1}^{\mu+1}{\left(1-r^2/a^2\right)^k\over k}
+2\ln\left({b\over a}\right)\right].
\eeq
The restriction of $\mu$ to an integer is of course not necessary, but 
gives simpler analytic expressions.

The expression in square brackets can be identified as $g$.  For the 
uniform distribution ($\mu$=0) on axis ($r$=0), we recover
\beq
g=1+2\ln (b/a).
\eeq
Generalizing to $g=K_0+2\ln (b/a)$, we find 
$K_0=\sum_1^{\mu+1}{1\over k}$.  See Table 1.

For the gaussian, the sum diverges logarithmically, but for constant 
$\tilde{x}$, $\ln (b/a)$ diverges as well and the two infinities 
cancel.\footnote{Actually, this violates our assumption $a<b$.  However,
in realistic cases $b>4\sigma_x$ so there is so little beam outside
$r=b$ that the approximation is not significant.} Defining $\tilde{K}_0$
by $g=\tilde{K}_0+2\ln (b/\tilde{x})$, we get 
\beq
\tilde{K}_0=K_0-\ln(\mu/2+1).
\eeq
$\tilde{K}_0(\mu)$ is also given in Table 1.  For large $\mu$, $K_0$ 
approaches $\gamma+\ln(\mu+3/2)$, where $\gamma$ is Euler's constant 
$0.57721566\cdots$.  Hence, for the gaussian, $\tilde{K}_0=\gamma+\ln 2$.
\begin{table}
\begin{center}
\caption{On-axis values for $K$}
\begin{tabular}{c||c|c}
$\mu$   & $K_0$  &$\tilde{K}_0$\\ \hline\hline
0       &1.0000  &1.0000 \\
1       &1.5000  &1.0945 \\
2       &1.8333  &1.1402 \\
3       &2.0833  &1.1670 \\
5       &2.4500  &1.1972 \\
10      &3.0199  &1.2281 \\
$\infty$&$\infty$&1.2704 \\
\end{tabular}
\end{center}
\end{table}

\section{The Average Potential}
One can find an average by first averaging the betatron motion for a 
given particle and then averaging over all the betatron amplitudes in 
the beam.  However, for a stationary transverse distribution, one can 
find the result more easily by simply averaging the potential over all 
particles in the beam at a given instant in time without regard to
betatron amplitudes.  In this way the average potential is
\beq
\la V\ra={\int V\rho rdrd\theta\over\int\rho rdrd\theta}=
{2\pi\over\lambda}\int_0^aV\rho rdr
\eeq
Integrating term by term we find
\beq
\la V\ra={-\lambda\over 4\pi\epsilon_0}
\left[\sum_{k=1}^{\mu+1}\left({1\over k}-{1\over k+\mu+1}\right)
+2\ln\left({b\over a}\right)\right].
\eeq
The summation can be identified as $\la K\ra$.  Again, it diverges as 
$\mu\rightarrow\infty$, but remains finite if the beam size is 
characterized by $\tilde{x}$ instead of $a$.  See Table 2.
\begin{table}
\begin{center}
\caption{Average values for $K$}
\begin{tabular}{c||c|c}
$\mu$   &$\la K\ra$&$\la\tilde{K}\ra$\\ \hline\hline
0       &0.5000  &0.5000 \\
1       &0.9167  &0.5112 \\
2       &1.2167  &0.5235 \\
3       &1.4488  &0.5325 \\
5       &1.7968  &0.5440 \\
10      &2.3489  &0.5572 \\
$\infty$&$\infty$&0.5772 \\
\end{tabular}
\end{center}
\end{table}
For those interested in special functions, the solution for general 
$\mu$ is
\beq
\la\tilde{K}\ra=\gamma+2\psi(\beta)-\psi(2\beta-1)-\ln(\beta/2),
\eeq
where $\beta=\mu+2$ and $\psi$ is the psi (or digamma) function.  An 
approximation that is within 0.2\% for $\mu\geq 1$ is 
$\la\tilde{K}\ra\approx\gamma-1/(4\mu+11)$.

\section{Conclusion}
The value of $\tilde{K}$ averaged over the beam is remarkably stable; 
varying by only 15\% over all the binomial distributions from K-V to 
gaussian.  It is therefore safe to conclude that for reasonably
`good' transverse distributions (i.e.\ smooth, peaked at the centre,
no halo), the form factor $g$ which appears in the longitudinal space
charge force and in the space charge impedance is given to a good
approximation by 
\beq
g={1\over 2}+2\ln \left({b\over\tilde{x}}\right)
\eeq
where $b$ is the beam pipe radius and $\tilde{x}$ is twice the rms beam 
size.  This should be used in place of $g=1+2\ln (b/a)$ for the 
calculation of space charge effects when longitudinal motion is 
considered independently of transverse motion.

\end{document}